\documentclass{article}

\usepackage{arxiv}

\usepackage[utf8]{inputenc} 
\usepackage[T1]{fontenc}    
\usepackage{hyperref}       
\usepackage{url}            
\usepackage{booktabs}       
\usepackage{amsfonts}       
\usepackage{amsmath,amssymb}
\usepackage{nicefrac}       
\usepackage{microtype}      
\usepackage{lipsum}
\usepackage{graphicx}
\usepackage{multirow}
\usepackage{float}
\graphicspath{ {./images/} }

\title{Creating Synthetic Datasets for Collaborative Filtering Recommender Systems using Generative Adversarial Networks}

\author{
 Jes\'{u}s Bobadilla \\
 Departamento de Sistemas Informáticos\\ 
 ETSI Sistemas Informáticos\\
 Universidad Politécnica de Madrid, C/ Alan Turing s/n\\
 28031, Madrid, Spain\\
  \texttt{jesus.bobadilla@upm.es} \\
   \And
 Abraham Guti\'errez \\
  Departamento de Sistemas Informáticos\\ 
 ETSI Sistemas Informáticos\\ 
 Universidad Politécnica de Madrid, C/ Alan Turing s/n\\
 28031, Madrid, Spain\\
  \texttt{abraham.gutierrez@upm.es} \\
  \And
 Raciel Yera \\
   Departamento de Inform\'{a}tica\\ 
 Universidad de Ja\'{e}n\\
 Ja\'{e}n, Spain\\
 Universidad de Ciego de \'{A}vila\\
 Ciego de \'{A}vila, Cuba\\
  \texttt{ryera@ujaen.es} \\
    \And
 Luis Mart\'{i}nez \\
   Departamento de Inform\'{a}tica\\ 
 Universidad de Ja\'{e}n\\
 Ja\'{e}n, Spain\\
  \texttt{martin@ujaen.es} \\
  }
 
\begin{document}
\maketitle
\begin{abstract}
Research and education in machine learning needs diverse, representative, and open datasets that contain sufficient samples to handle the necessary training, validation, and testing tasks. Currently, the Recommender Systems area includes a large number of subfields in which accuracy and beyond accuracy quality measures are continuously improved. To feed this research variety, it is necessary and convenient to reinforce the existing datasets with synthetic ones. This paper proposes a Generative Adversarial Network (GAN)-based method to generate collaborative filtering datasets in a parameterized way, by selecting their preferred number of users, items, samples, and stochastic variability. This parameterization cannot be made using regular GANs. Our GAN model is fed with dense, short, and continuous embedding representations of items and users, instead of sparse, large, and discrete vectors, to make an accurate and quick learning, compared to the traditional approach based on large and sparse input vectors. The proposed architecture includes a DeepMF model to extract the dense user and item embeddings, as well as a clustering process to convert from the dense GAN generated samples to the discrete and sparse ones, necessary to create each required synthetic dataset. The results of three different source datasets show adequate distributions and expected quality values and evolutions on the generated datasets compared to the source ones. Synthetic datasets and source codes are available to researchers.  
\end{abstract}


\section{Introduction}
\label{sec:introduction}

Recommender systems (RS) are a relevant area in artificial intelligence due to the growing popularity of social networks. The big companies that extensively use RS are TripAdvisor, Netflix, Spotify, YouTube Music, TikTok, You Tube and Amazon \cite{Fang2016}. These companies make use of the RS models to recommend users similar items (music, videos, trips, news) to those that they have already consumed; some other companies, such as Facebook, work hard collecting their client activity to provide personalized publicity instead personalized products or services. RSs are usually classified attending to their filtering approach \cite{Yera2017}; content-based RSs select the recommended items looking for similar contents \cite{Yera2022b}; since most of the item contents are texts, natural language processing models are used. Reviews \cite{Zheng2017} and tweets \cite{Gong2016} are two usual types of content-based filtered data. Product images can also be processed to make recommendations; convolutional neural networks are the most used models to perform this task \cite{Kanwal2020}. Social filtering has been extensively used to improve social-based recommendations. This type of filtering uses data such as tags, followers, and followed, and makes use of the concepts of reputation and trust \cite{McNally2014}. Geographic information such as GPS coordinates and POI is used mainly to support context-aware filtering \cite{Villegas2018}. Demographic filtering (age, gender, country, etc.) has been usually combined with other filtering types, implementing recommendation ensembles \cite{Moradi2019}. Beyond the mentioned filtering strategies, collaborative filtering (CF) \cite{Jalili2018} is the most important approach to implement RSs, since it provides superior accuracy, particularly when combined with some other filtering types. Effective RS research makes use of innovative models, adequate quality measures, and representative datasets. Whereas RS research is mainly focused on models and measures, this paper tries to make progress in CF datasets.

The historical evolution of CF begins with the use of memory-based models, mainly the K-Nearest Neighbors algorithm \cite{Zhu2018}. Memory-based approaches were replaced by model-based machine learning ones due to their overall performance: they are superior in the accuracy of the results, also in the time to obtain predictions (once the model has learnt); and their output able to be explained through post-hoc techniques \cite{Yera2022a}.  Matrix Factorization (MF) \cite{D2022} has been the most used machine learning model to implement collaborative filtering; it makes a dimensional reduction of users and items, catching the main patterns that relate them with the casted votes. Additionally, by using Non-Negative Matrix Factorization (NFM) \cite{Aghdam2022}, semantic meanings can be assigned to latent factors. Bayesian NMF \cite{Ayci2019} makes it possible to simultaneously group users and make predictions, opening the doors to the effective recommendation to group of users and the social clustering applications \cite{Bobadilla2017}. Currently, CF research is based on deep learning models, where DeepMF \cite{Xue2017} was the basis for modern approaches. DeepMF is the model that we use in this paper, in which users are coded in a latent space by means of an embedding layer, whereas items are coded in another different latent space by means of a second embedding layer; finally, predictions are made by making the dot product of both, the item and the user, embeddings. DeepMF improves MF due to the inherent competence of neural networks to catch the non-linear relations between samples. Neural Collaborative Filtering (NCF) \cite{He2017} is extensively used to implement CF; this model replaces the DeepMF dot layer by a Multi-Layer Perceptron (MLP) and it outperforms to DeepMF when applied to large and complex datasets. Beyond accuracy, deep learning models are emerging to perform some innovative tasks, such as improving fairness, where the DeepFair model \cite{Bobadilla2021} gets a trade-off between equity and precision; green computing \cite{Himeur2021}; results explanation via latent space visualization \cite{Bobadilla2022a} and efficient neighborhood identification \cite{bobadilla2022b}. The adversarial network-based recommendation has recently been introduced in the RS area \cite{Zhang2019deep} and we focus on it in the 'Related work' section.  Generative Adversarial Networks (GAN) \cite{Goodfellow2020} are responsible for popular fake faces and fake videos that flood social networks. Their architecture has two separated neural networks that compete one against the other ('adversarial'), such as an art counterfeiter competing against an art expert, ensuring that both improve their work. The GAN 'counterfeiter' is a generator model that creates fake samples from random noise vectors, while the GAN 'expert' is a discriminator model implemented as a simple binary classifier: fake, non-fake.

Quality measures are a key piece to make effective research, since they make it possible to compare state-of-the-art baselines with proposed algorithms, methods, and models. Beyond the usual prediction and recommendation quality measures (MAE, MSD, precision, recall, F1, NDCG, etc.), some other measures, such as novelty and diversity \cite{Sacharidis2019}, have recently acquired growing importance. From those, diversity focuses the current attention of researchers, due to the risks of inadequate recommendations in social networks, such as those which exhibit lack of variability and promote prefixed ideas and behaviors. Diversity and reliability in RS have been improved by introducing diversity-enhancing constraints in the MF model \cite{Gogna2017}; additionally, a deep learning classification model \cite{Bobadilla2022c} is proposed to obtain the recommendation reliability values from the softmax output layer of the neural network. Quality values are obtained when a model or method is tested on balanced CF datasets. To obtain balanced training and testing sets, referred to their distributions of users and items, deterministic strategies are proposed in \cite{Pajuelo2019}. Most of the RS research makes use of popular CF datasets such as Movielens, FilmTrust, MyAnimeList or CiteSeer; CF datasets include different domains such as music, movies, POIs, tourism, news, research papers, tagged data, etc. Some of these datasets have been filled with explicit votes from users, while others contain implicit interactions between users and systems. There are also datasets filled with crawled Web pages or academic PDFs \cite{Bollacker1998} and some others are enriched with social tags that researchers add to the articles \cite{Choochaiwattana2010}. A selection of relevant social CF datasets \cite{Shokeen2020} is provided and related to some articles using them. Recently, an educational news dataset \cite{Xing2020} was released; it included contextualized information: time and location. Finally, an RS dataset is provided that contains artificial intelligence research data \cite{Ortega2018a} to obtain segmented information, clustering, and geographical locations. It is particularly relevant that no synthetic datasets have been used yet, consequently the CF research does not benefit from the flexibility that parameterization provides in the design of experiments: different dataset sizes, number of users and items, and so on; and this paper aims at filling the gap.

The rest of the paper has been structured as follows: in Section 2 the related work is introduced, focusing on the most recent uses of the GAN models applied to RS. Section 3 explains the proposed model and its formalization. Section 4 shows the design, result, and discussion of the experiments. Finally, Section 5 contains the main conclusions of the article and future work.

\section{Related works}

Generative deep learning is an innovative field in the CF RS area. Although some variational autoencoder approaches have been published \cite{Liang2018_1, Zamany2022_2}, current research has focused on GAN models \cite{Gao2021_3}. A CF subfield where GANs are used is attack/defense strategies \cite{Deldjoo2021_4}, where these models can reinforce the security in RS. Nevertheless, the most extended uses of CF GANs are: a) to solve the data noisy issue, and b) to tackle the data sparsity problem, and implement a data augmentation framework by capturing the distribution of real data. CFGAN \cite{Chae2018_5} is a model that generates vectors of purchases rather than IDs of items, and then uses the generated fake purchase vectors to augment the real ones. The Wasserstein version of CFGAN is the unified GAN (UGAN) \cite{Wang2019} and reports improvements compared to CFGAN. To prioritize long and short-term RS information (interactions between users and items that change quickly or slowly), the PLASTIC \cite{Zhao2018_6} model trains a generator and uses it as a reinforcement learning agent. The recurrent GAN: RecGAN \cite{Bharadhwaj2018_7}, learns temporal patterns in ratings; they combine GAN and recurrent neural networks models. To capture negative sampling information in the CF datasets, IPGAN \cite{Guo2020_8} implements two different generative models: one for positive instances and one another for negative instances. IPGAN considers the relations between the positive ratings sampled and the negative ones selected.

Currently, the DCGAN model \cite{Zhao2022_9} combines GAN and reinforcement learning models to catch the information of the RS sessions, rather than the traditional historical matrices of votes from users to items. Session information includes the responses of users to current recommendations. The user’s immediate feedback is managed by the reinforcement learning model combined with the GAN. The NCGAN \cite{Sun2022_10} incorporates a neural network to extract nonlinear features from users, and a GAN to guide the recommendation training; the generator model makes user recommendations, whereas the discriminator model measures distances between real and generated distributions. An innovative method to improve the information flow from generator to discriminator \cite{Lin2021} reduces the discrepancies between both models in the CF GAN. A regularization Wasserstein GAN model is used in \cite{Wang2021}, combined with an autoencoder acting as a generator, reporting accuracy improvement when applied to high-dimensional and sparse CF matrices. A CGAN (Conditional GAN) is used \cite{Wen2022} to improve CF recommendations, and the sizes of the rating vectors can be set, simplifying the generator and discriminator tasks. Additionally, it allows to stablish conditional rating generation. For datasets that do not follow standard Gaussian distributions, a missing data imputation based on GAN \cite{Deng2022} is proposed; results show a quality improvement on several representative classification data sets. Trust information is used in \cite{Chen2021} to make effective recommendations. They propose a GAN where the discriminator is an MLP model, and the generator is an LSTM model. Finally, CF datasets are usually imbalanced due to their social data collection (e.g: more young than old people). To address this limitation, \cite{Shafqat2022} proposes a Wasserstein GAN model in the generator, and the PacGAN concept in the discriminator \cite{Lin2018}, to minimize the mode collapse problem.

A platform for multi-agent RS simulation is the probabilistic-based RecSim \cite{Mladenov2020}, that generates synthetic profiles of users and items, and it uses Markov chains and recurrent neural networks. The Virtual-Taobao \cite{Shi2019} is a multiagent reinforcement learning system designed to improve search in the social Taobao web site; it makes use of a GAN for simulating internal distributions. A simple matrix factorization is used \cite{Ziegler2005} to inject topic diversification in the recommendation process. The DataGenCars \cite{DelCarmen2017} is a Java-based generator of RS synthetic data; it contains a statistical basement that provides flexibility, but it returns low accuracy compared to deep learning generative models. Finally, the SynEvaRec framework \cite{Provalov2021} provides synthetic RS datasets generation using the Synthetic Data Vault (SVD) library. This library models multivariate distributions using copula functions; its CTGAN sub-library includes GAN models. The main advantage of SynEvaRec, compared to previous frameworks, is that it can take different RS as source; its main drawbacks are the poor quality of the results in most of the cases, and the excessive time consumed to accomplish the training stage. 

Previous research mainly focuses on improving different objectives such as noise reduction, recommendation quality, prediction values, defense against attacks, or balancing data. To make this happen, many different approaches and information sources have been combined: use of GAN, CGAN, Wasserstein GAN, etc. GAN models have been combined with RNN and LSTM temporal networks, and reinforcement learning has been introduced in the GAN based architectures. Long and short data has been introduced to the proposed models, in addition to trust information, session logs, including response of the users to previous recommendations, and inferred negative votes. The pure generation of synthetic datasets does not seem to be an objective in this novel field of GAN applied to CF RS, currently focused on improving prediction and recommendation quality results by means of data augmentation based on the inherent GAN model ability to catch the nonlinear complex patterns of the high-dimensional and sparse CF datasets. The innovation of our proposal is to generate representative and useful CF synthetic datasets, rather than improving different quality results on existing ones. Additionally, it allows to set representative parameters and to obtain a whole ‘family’ of synthetic datasets, taking real datasets as source such as Movielens, FilmTrust, or MyAnimeList. These parameters are the number of users, the number of items, the number of samples and the variability of the generated data. By varying the parameter values, we can generate different versions of the same CF pattern, such as a Movielens-based dataset containing 8000 users and 3000 items, another Movielens-based dataset containing 2000 users and 1000 items, and so on. In this way we can test the accuracy and performance impact of the dataset size, its sparsity, its number of users and items, e.g.: the MAE improvement as the number of users increases. As far as we know, there are no published methods or models to create, in a parametrized deep learning model, accurate and scalable synthetic datasets from diverse sources.

\section{The Generative Adversarial Networks-based approach for datasets building in collaborative filtering}

This section proposes the GANRS method, that makes use of a GAN network to generate synthetic CF datasets; the GAN is fed with a real CF dataset and the model learns its internal patterns. The most innovative contribution is to feed the GAN with dense and small embedding representations of users and items, instead of the traditional approach where the GAN inputs are large and sparse vectors containing the votes casted for each user. The main advantage of the GANRS method is to greatly reduce the complexity of the GAN architecture, its convergence speed, and its performance.

\begin{figure}
     \centering
     \includegraphics[width=1\textwidth]{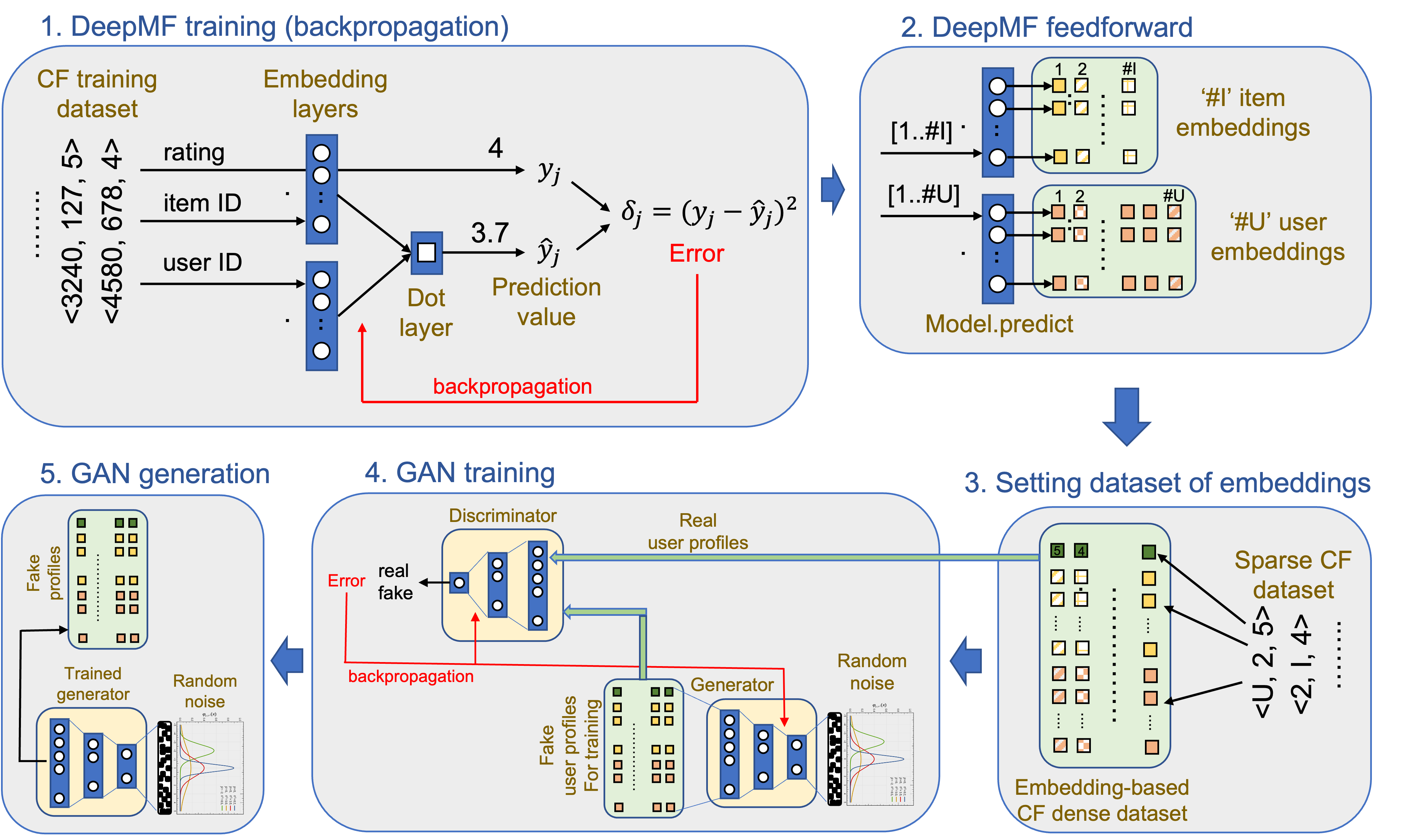}
     \caption{Main stages in the GANRS method: generation of dense fake samples.}
     \label{fig:fig1}
\end{figure}

To illustrate the GANRS method, we will use Figures \ref{fig:fig1} and \ref{fig:fig2}, which are also linked to the below formalization presented at the end of this section across the step by step process enumerated. Figures \ref{fig:fig1} and \ref{fig:fig2} show the seven designed stages to generate different synthetic datasets from real datasets (Movielens, Netflix, etc.). Stage 1 (top left graph in Figure \ref{fig:fig1}) shows the training of a DeepMF model used to set both the embedding layer of users and the embedding layer of items. Basically, embedding layers on a neural network efficiently convert an input from a sparse representation into an output dense representation. For each input sample $<user, item, rating>$ in the training set, the output dot layer combines the embedding layer values to predict the rating value and to obtain the output error "(rating - prediction)" that will we backpropagated to update the learning parameters. Steps 6 to 10 formalize these concepts. Once the DeepMF model has learned, stage 2 (top right graph in Figure \ref{fig:fig1}) shows the DeepMF feedforward process where each item ID (from 1 to the number or items in the dataset) feeds the item embedding, which outputs the item ID dense representation; usually, CF embedding vectors have a size from 5 to 10. Same applies to user IDs as input and their output dense representations.  Please note that the number of items in the dataset will be different from the number of users. Steps 11 to 14 explain this second stage.

The purpose of the third stage is to convert the source sparse CF dataset into its dense representation. To accomplish the task, for each source $<user, item, rating>$ sample in the dataset (e.g.: $<8920, 345, 4>$) we replace the user ID (8920 in this example) for its related dense representation; same applies for the item ID. Using embeddings of size 5, the result in the example could be such as: $$<[0.03, 0.94, 1.02, 0.87,-0.78], [-1.23, 0.99, 1.02, 0.65, -0.48], 4>$$ Stage 3 in Figure \ref{fig:fig1} shows an illustrative example. Step 15 formalizes the operative. The dense dataset obtained will be used in stage 4 to train a GAN capable of generating fake profiles. Our GAN will use the stage 3 dense dataset to train the discriminator by providing it with the necessary real samples. The GAN generator takes Gaussian random noise as input and iteratively learns how to generate increasingly good fake profiles capable of cheating the discriminator model. Once the generator and the discriminator have learnt, the generator can convert input noise vectors into dense samples that mimic the patterns of the real dataset provided in stage 3. Stage 4 is formalized in steps 16 to 18.

The last stage in Figure \ref{fig:fig1} (bottom left graph) makes use of the trained generator model of the GAN (stage 4) to generate as many fake samples as desired. We feed the generator with successive vectors of random noise values following a Gaussian distribution, and the generator outputs successive fake dense samples following the patterns of the real dataset (obtained in stage 3). The higher the standard deviation of the Gaussian distribution, the higher the variety of individual values in the generated dense fake samples. As an example, a low standard deviation value in the random noise Gaussian distribution leads to a higher proportion of votes ‘3’ (in the range of 1 to 5), while choosing a high standard deviation value will produce a higher density of votes ‘5’ and ‘1’. Step 19 formalizes the generation of fake samples. 

Even though it could be considered that the GANRS method is completed, it is not the case because our goal is to generate fake datasets of sparse samples (such as Movielens or FilmTrust); then it is necessary to convert from the obtained dense representation in stage 5 to the usual sparse representation seen in stage 1. The process is not straightforward, since all the dense representations of the fake samples are different from each other; this will be better explained using the example in Table \ref{tab:samplesrepresentation}: it can be observed that the user 890 (two first rows) has very similar dense embedding values, but there are not identical, since the GAN generator is not able to create the same exact values from the noise input vectors. The same situation occurs in Table \ref{tab:samplesrepresentation} for the item with ID 31. Consequently, the GANRS method provides the way to 'group' similar dense embeddings into a unique ID; that is, to convert the dense bold vectors of the user in Table \ref{tab:samplesrepresentation} into a unique user ID (it is not necessary to be 890), and the dense bold vectors of the item into a unique item ID (it is not necessary to be 31, neither).

\begin{table}[H]
\begin{footnotesize}
\begin{center}
\begin{tabular}{cc}
\hline
        Sparse&Dense\\
\hline

$<\textbf{890}, 47, 5>$&$< \textbf{[0.03, 0.94, 1.02, 0.87,-0.78]}, [-1.23, 0.99, 1.02, 0.65, -0.48], 5 >$ \\
$<\textbf{890}, \textbf{31}, 4>$&$< \textbf{[0.02, 0.95, 0.99, 0.81,-0.69]}, \textbf{[0.45, -0.78, 0.83, -0.15, 0.09]}, 4 >$ \\
$<968, \textbf{31}, 4>$&$< [-1.04, 0.04, 0.66,-0.67,0.11], \textbf{[0.42, -0.71, 0.80, -0.10, 0.14]}, 4 >$ \\
$<123, \textbf{31}, 2>$&$< [1.56, -1.12, 0.33,1.22,-0.87], \textbf{[0.43, -0.75, 0.80, -0.11, 0.06]}, 2 >$ \\

\hline
\end{tabular}
\caption{Example of samples representation}
\label{tab:samplesrepresentation}
\end{center}
\end{footnotesize}
\end{table}

\begin{figure}
     \centering
     \includegraphics[width=0.5\textwidth]{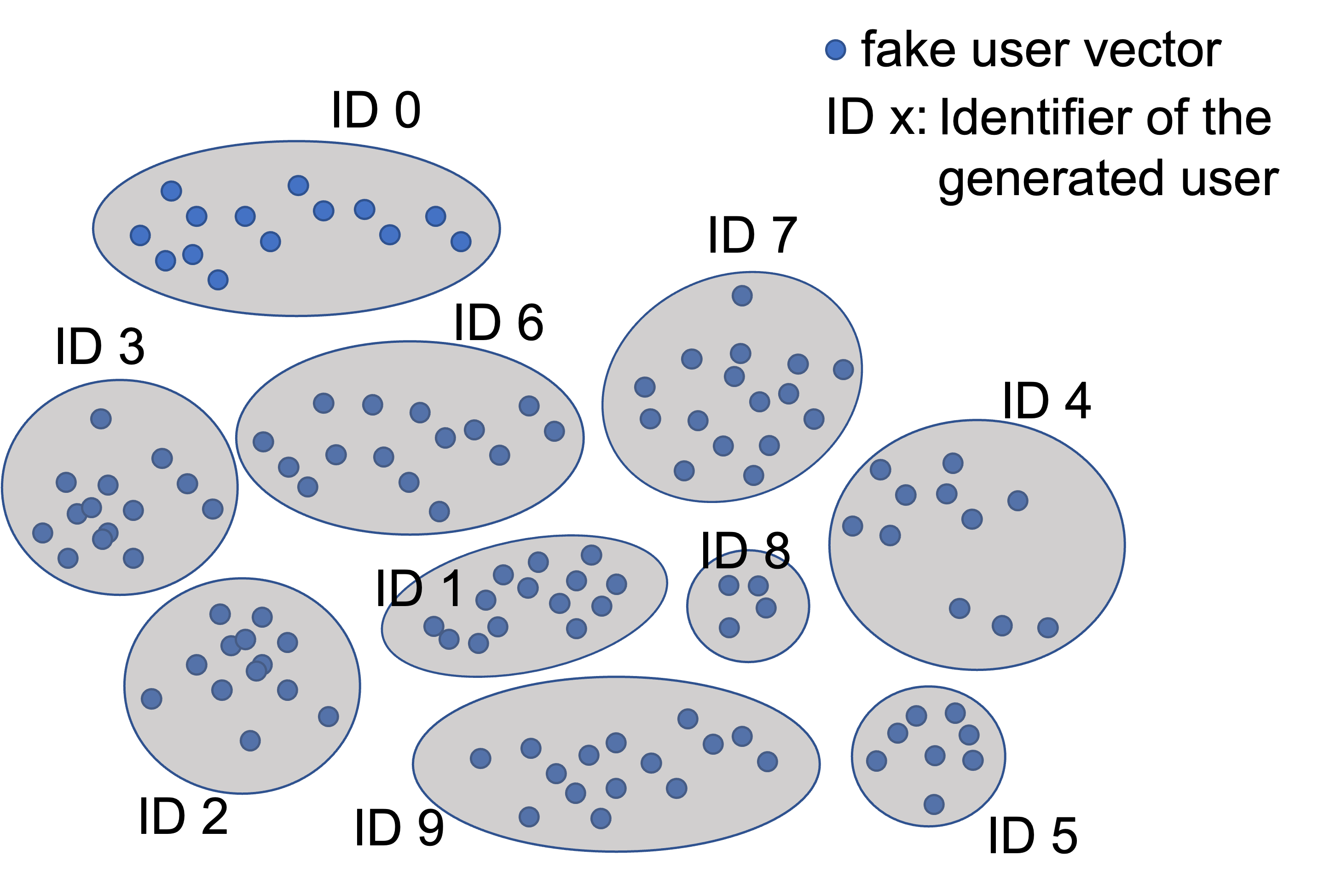}
     \caption{Grouping fake user vectors (blue circles) and the assignation of a fake user ID to each one of them (ID x). Gray ellipses represent k-means clustering groups.}
     \label{fig:figx}
\end{figure}

\begin{figure}
     \centering
     \includegraphics[width=1\textwidth]{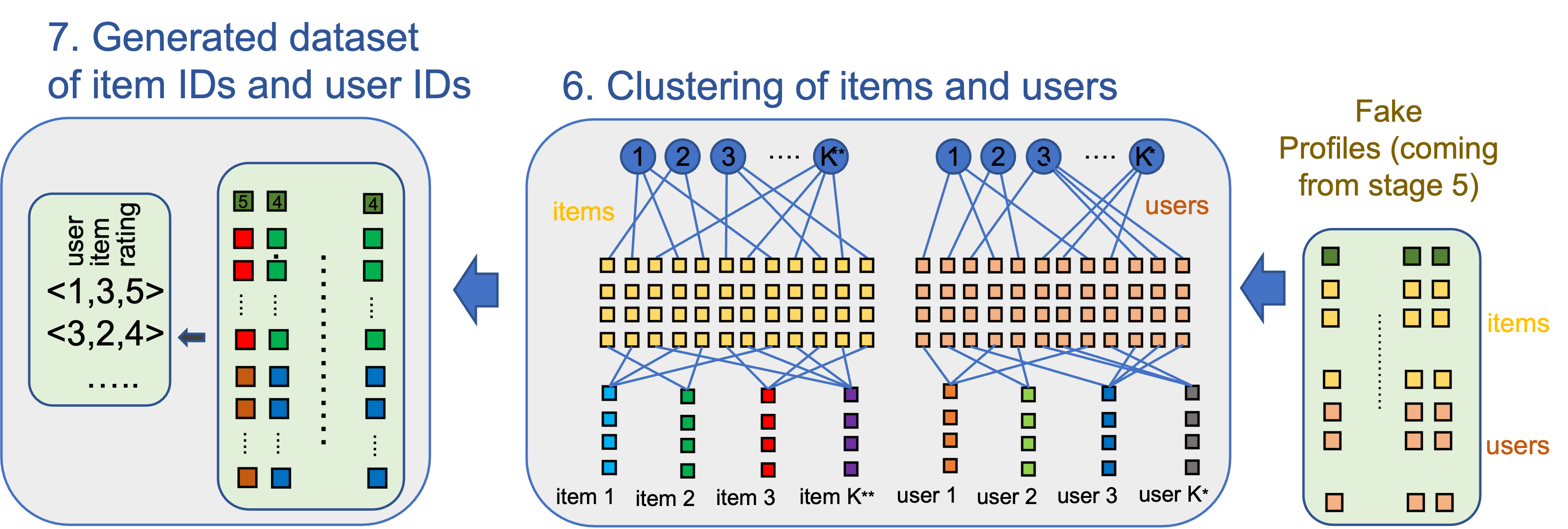}
     \caption{Final stages in the proposed GANRS method: conversion of dense samples into sparse samples.}
     \label{fig:fig2}
\end{figure}

To group similar dense embeddings into a unique ID, a K-Means clustering \cite{Ahmed2020k} has been chosen. This algorithm has the relevant property that a number K of clusters must be chosen a priori, and it is very convenient in this context, since, in this way, we will have the opportunity to establish the number of users and the number of items in the GANRS synthetic generated dataset. Stage 6 of Figure \ref{fig:fig2} shows the concept, where $K^*$  has been selected as number of users and $K^{**}$ has been selected as number of items. Two separated K-Means processes are run: one to group user embeddings, and the other to group item embeddings. Steps 20 to 23 formalize these two clustering processes. To better understand this stage, we can consider an example where one million fake samples have been generated and we want to create a synthetic dataset containing two thousand fake users and one thousand fake items. To accomplish the task (shown in Table \ref{tab:samplesrepresentation}), we should obtain two thousand groups collected from the one million user vectors (same for the one thousand groups of items). On average, five hundred user vectors could be assigned to each user group (and, analogously, one thousand item vectors to each item group), but we know that it depends on the user and item vector patterns. To adequately accomplish the grouping task, machine learning provides us with the clustering algorithms, of which the k-means allow us to set the number of desired groups (two thousand for users and one thousand for items, in our example). Running both clustering processes (one for users and the other for items) we can assign a fake user ID to all the fake user vectors in each cluster. Please note that the ID number can be assigned at random to each of the two thousand clusters (same for the one thousand IDs of the items). Figure \ref{fig:figx} illustrates the concept, where the blue circles represent the one million fake vectors in our example, and the "ID number" text represents the two thousand required users. Gray ellipses are the result of the K-means clustering. As a result: a) each group of 'similar', 'nearest', 'more related' user vectors shares the same fake user ID, b) all user vectors have an assigned fake user ID, and c) each fake user is related to a potentially different number of user vectors.  

The seventh stage in Figure \ref{fig:fig2} converts dense fake samples (coming from stage 5) into sparse samples $<user, item, rating>$. To accomplish this task, for each sample in the dense representation we replace its user vector by its centroid number (from 1 to $K^*$) and its item vector by its centroid number (from 1 to $K^{**}$); the rating value remains the same. Figure \ref{fig:fig2} shows an example of this operative, formalized in the step 24. Please note that repeated samples will appear in the previous discretization process, since the GAN generator can create very similar dense samples that will be converted to the same discrete encoding. There are several factors that modulate the number of repeated samples, such as the number of generated samples, the embedding size, the size of the noise vector and the standard deviation of the Gaussian distribution, but the most relevant factor is the number of chosen users or items ($K^*$ and $K^{**}$): the lower the $K$, the higher the number of repeated samples. When the number of users or items is low, the average number of samples grouped in each cluster is high. Step 25 formalizes the process of removing repeated discrete samples. Finally, the GANRS method can generate a small proportion of samples in which different votes are cast from the same user to the same item; e.g.: $<879, 56, 4>$, $<879, 56, 5>$. This could be considered as a convenient behavior: code a higher range of votes (4.5 in the example) or express a change in the user's opinion. These cases can be unchanged, changed, or removed. Step 26 formalizes their removal operation.

Appendix A (Table \ref{tab:appa}) shows the main parameter and hyperparameter values used to design both the models DeepMF and GAN involved in the proposed GANRS method. More detailed information can be found by analyzing the provided source code available at \url{http://suleiman.ujaen.es:8061/gitlab-instance-981c80cc/ganrs}. The formalization of the GANRS method is presented and structured according to the seven stages explained in this section and shown in Figures \ref{fig:fig1} and \ref{fig:fig2}:

\begin{itemize}
\item Stage 0. CF definitions
\begin{itemize}
\item [1] \label{st:1} Let $U$ be the set of users who make use of a CF RS. 
\item [2] Let $I$ be the set of items available to vote on in the CF RS. \label{st:2}
\item [3] Let $V$ be the range of allowed votes; usually $V=\{1,2,3,4,5\}$.\label{st:3}
\item [4] Let $S$ be the set of samples contained in the CF dataset; where $N = |S| = the total number of cast votes$.\label{st:4}
\item [5] $S=\{<u,i,v>_1, <u,i,v>_2,….,<u,i,v>_N\};$ where each $u\in\{1,…,|U|\}$, each $i\in\{1,…,|I|\}$, and each $v\in\{1,…,|V|\}.$\label{st:5}
\end{itemize}
\item Stage 1. DeepMF training
\begin{itemize}
\item [6]\label{uno} Let E be the size of two neural layer embeddings used to vectorize each user and each item belonging to U and I, respectively.
\item [7]\label{st:7} Let $f^{eu}(u)=[e_0^u,e_1^u,…,e_E^u]$, where $f^{eu}$ is the embedding layer output of the users, where $u\in \{1,…,|U|\}$.
\item [8] \label{st:8} Let $f^{ei}(i)=[e_0^i,e_1^i,…,e_E^i]$, where $f^{ei}$ is the embedding layer output of the items, where $i\in \{1,…,|I|\}$. 
By combining both dense vectors of user and item embeddings: ($[e_0^u,e_1^u,…,e_E^u]$ and $[e_0^i,e_1^i,…,e_E^i])$, we can make rating predictions in the DeepMF training stage.
The dot product of the user embedding and the item embedding in each $<u,i,v>_j \in S$ provides its rating prediction:
\item [9]\label{st:9} $\hat{y}_j= f^{eu}(u) \cdot f^{ei}(i)= [e_0^u,e_1^u,…,e_E^u]\cdot[e_0^i,e_1^i,…,e_E^i]$
\item [10]\label{st:10} $\frac{1}{2}(y_j-\hat{y}_j)^2$ is the output error used in the DeepMF neural network to start the backpropagation algorithm, where the neural weights are iteratively improved from the $\delta_j$ values: $\triangle w_{ji}= \alpha y_j f'(Net_i) \sum_k w_{ik}\delta_k$, when $k$ is a hidden layer, and $\triangle w_{ji}= \alpha y_i f'(Net_i) \frac{1}{2}(y_k-\hat{y}_k)^2$, if $k$ is the output layer. i, j, and k are successive sequential layers.
\end{itemize}
\item Stage 2. DeepFM feedforward

Once the DeepMF has learned, we can collect the embedding representation of each user and each item in the CF RS.
\begin{itemize}
\item [11] Let $E^*=\{<u,[e_0^u,e_1^u,…,e_E^u ]>,\forall u \in U\}$, be the set of embeddings for all the RS users.\label{st:11}
\item [12] Let $E^*(u) = [e_0^u,e_1^u,…,e_E^u ]$\label{st:12}
\item [13] Let $E^{**}=\{<i,[e_0^i,e_1^i,…,e_E^i ]>,\forall i \in I\}$, be the set of embeddings for all the RS items.\label{st:13}
\item [14] Let $E^{**}(i) = [e_0^i,e_1^i,…,e_E^i ]$\label{st:14}
\end{itemize}
\item Stage 3. Setting the dataset of embeddings
\begin{itemize}
\item [15] Let $R=[<E^*(u),E^{**}(i),v>],\forall <u,i,v>_j\in S$ be the embedding-based dataset of real samples.\label{st:15}
\end{itemize}
\item Stage 4. GAN training
\begin{itemize}
\item [16] Let $f^D$ be the discriminator D model belonging to a GAN model.\label{st:16}
\item [17] Let $f^G$ be the generator G model belonging to a GAN model.\label{st:17}
\item [18] Let $f^{GD}$ be the optimization function of the GAN model;\label{st:18}
$f^{GD}=Min_GMax_D f(D,G)=E_R [log(D(R))]+E_z [log(1-D(G(z)))]$, where $E_R$ is the expected value for real samples, $z$ is the random noise that feeds the generator $G$, and $E_z$ is the expected value for the generated fake profiles $G(z)$. Note that $R$ refers to [15].
\end{itemize}
\item Stage 5. GAN generation
\begin{itemize}
    \item [19] Let $F=f^G (z)$ be the generated dataset of fake samples from different random noise vectors $z$.\label{st:19}
\end{itemize}
\item Stage 6. Clustering of items and users.
\begin{itemize}
    \item [20] Let $K^*$ be the number of clusters used to group the embeddings of the users.\label{st:20}
    \item [21] Let $K^{**}$ be the number of clusters used to group the embeddings of the items.\label{st:21}
    \item [22] Let $h^*(u)=c | c \in \{1,…,K^*\}$, be the clustering operation that assigns a centroid to each user.\label{st:22}
    \item [23] Let $h^{**} (i)=c | c \in \{1,…,K^{**})\}$, be the clustering operation that assigns a centroid to each item.\label{st:23}
\end{itemize}
\item Stage 7. Setting dataset of item IDs and user IDs
\begin{itemize}
\item [24] Let H be the item IDs and users IDs discrete dataset obtained from the embedding-based dataset F of fake samples.  $H=\{<h^*(u),h^{**}(i),v> | \forall <E^*(u),E^{**}(i),v> \in F\}$\label{st:24}
\item [25] Let $S=\{H\}$ be the synthetic generated dataset version of H where duplicated samples are removed.\label{st:25}
\item [26] Let $G'=\{<h^*(u),h^{**}(i),v> \in H |  \nexists <h^* (u'),h^{**} (i'),v'> \in H$ where  $h^* (u)=h^*(u')  \land h^* (i)=h^{**}(i) \land v \neq v'\}$\label{st:26}
\end{itemize}
\end{itemize}

\section{Experiments and results}
Here it is evaluated the suitability of the presented procedure focused on building synthetic datasets. At first, it is presented the traditional datasets that will be used as starting point for the current procedure, as well as described the experiments to performed. Subsequently, the obtained results are showed and discussed.
\subsection{Experiments}
To test the proposed GANRS method behavior, we will use three representative and open datasets in the CF field: Movielens, Netflix and MyAnimeList. We have chosen the Movielens 100K version and a reduced version of the complete Netflix dataset: Netflix*, available in \cite{Ortega2018cf4j}. Table \ref{tab:parameters} shows the main parameter values for these datasets. A complete set of experiments has been run using Netflix*, whereas only a subset of these experiments is shown for Movielens and MyAnimeList, to reduce the paper size. Results testing Movielens and MyAnimeList are summarized at the end of this section. Each of the three source datasets is used to generate their corresponding synthetic versions: setting different numbers of users, items, and samples, and changing the standard deviation of the Gaussian random noise.

\begin{table}[H]
\begin{footnotesize}
\begin{center}
\begin{tabular}{cccccc}
\hline
        Dataset&\#users&\#items&\#ratings&scores&sparsity\\
\hline
Movielens 100K	&943	&1682	&99,831	&1 to 5&	93.71\\
\hline
Netflix*	&23,012	&1,750	&535,421	&1 to 5&	98.68\\
\hline
MyAnimeList	&19,179	&2,692	&548,967	&1 to 10	&98.94\\
\hline
\end{tabular}
\caption{Main parameter values of the tested datasets}
\label{tab:parameters}
\end{center}
\end{footnotesize}
\end{table}

Table \ref{tab:datasets} shows the GAN generated synthetic datasets used to test the proposed GANRS method, using Netflix* as source data. The grey cells show the number of generated datasets; 'std' is the standard deviation used in the random noise Gaussian distribution; \#users and \#items are the total number of users and items chosen to generate each dataset; \#samples is the number of fake samples created by the GAN generator. Please note that the final number of samples contained in each of the datasets is lower than \#samples, due to the removing process of repeated samples. Cases 1 to 15 in Table \ref{tab:datasets} are used to test the effect of changing standard deviation and number of users. Cases 16 and 17 test the consequences of increasing the number of items. Finally, cases 18, 19 and 20 test the behavior of the synthetic datasets when they have different sizes (number of samples). All generated datasets and the source code of the proposed GANRS method are fully available in \url{http://suleiman.ujaen.es:8061/gitlab-instance-981c80cc/ganrs}. Additionally, Appendix B (Figure \ref{fig:appb}) shows an example of the distribution graphs obtained for each of the synthetic datasets. Following the open access link   
\url{http://suleiman.ujaen.es:8061/gitlab-instance-981c80cc/ganrs}, each generated dataset is located on its specific directory where a 'readme.txt' file is provided along the synthetic dataset distribution graphs.

\begin{table}[H]
\begin{footnotesize}
\begin{center}
\begin{tabular}{|cccc|cccc|cccc|c|}
\hline
        \#	&std	&\#users	&\#items	&\#	&std	&\#users&	\#items	&\#	&std	&\#users	&\#items	&\#samples\\
\hline

1	&2.0	&100	&4000	&6	&2.5	&100	&4000	&11	&3.0	&100	&4000	&\multirow{6}{*}{1.5M}\\

2	&2.0	&1000	&4000	&7	&2.5	&1000	&4000	&12	&3.0	&1000	&4000	&\\

3	&2.0	&2000	&4000	&8	&2.5	&2000	&4000	&13	&3.0	&2000	&4000	&\\

4	&2.0	&4000	&4000	&9	&2.5	&4000	&4000	&14	&3.0	&4000	&4000	&\\

5	&2.0	&8000	&4000	&10	&2.5	&8000	&4000	&15	&3.0	&8000	&4000	&\\

16	&1.5	&4000	&2000	&17	&1.5	&4000	&8000	&&&&&\\	\hline			

18	&1.2	&2000	&4000	&&&&&&&&&								150K\\\hline

19	&1.2	&2000	&4000	&&&&&&&&&									500K\\\hline

20	&1.2	&2000	&4000	&&&&&&&&&									1M\\\hline

21	&1.2	&2000	&4000	&&&&&&&&&									3M\\

\hline
\end{tabular}
\caption{Parameter values of the synthetic datasets generated by GAN. Source: Netflix*. }
\label{tab:datasets}
\end{center}
\end{footnotesize}
\end{table}

Using the parameter values of Table \ref{tab:datasets}, a variety of experiments have been conducted. The classification of the experiments is as follows:

\begin{enumerate}
    \item Number of users
    \begin{enumerate}
    \item Distribution of users versus ratings
    \item Distribution of the user ratings
    \item Number of repeated samples
    \item Proportion of samples with the same user and item
    \item MAE and accuracy of the data set
    \item Users' precision and recall
    \end{enumerate}
    \item Number of items
    \begin{enumerate}
        \item MAE and accuracy of the dataset
        \item Item’s precision and recall
    \end{enumerate}
    \item Number of samples
    \begin{enumerate}
        \item Number of samples generated
        \item Precision and Recall
    \end{enumerate}
\end{enumerate}

\subsection{Results}

This subsection shows the obtained graphs when the designed experiments (previous subsection) are run. The synthetic datasets described in Table \ref{tab:datasets} are used to get results that allow us: 1) to compare the distributions of users, items and ratings belonging to the source datasets, in relation to the ones obtained using the synthetic ones, 2) to measure the quantity of repeated samples returned in the clustering stage, and 3) to test the prediction and recommendation qualities and trends obtained by running the proposed RSGAN method and comparing them to those shown by the source datasets. 

\subsubsection{Experiment 1a. Number of users: Distribution of users versus ratings}

\begin{figure}
     \centering
     \includegraphics[width=1\textwidth]{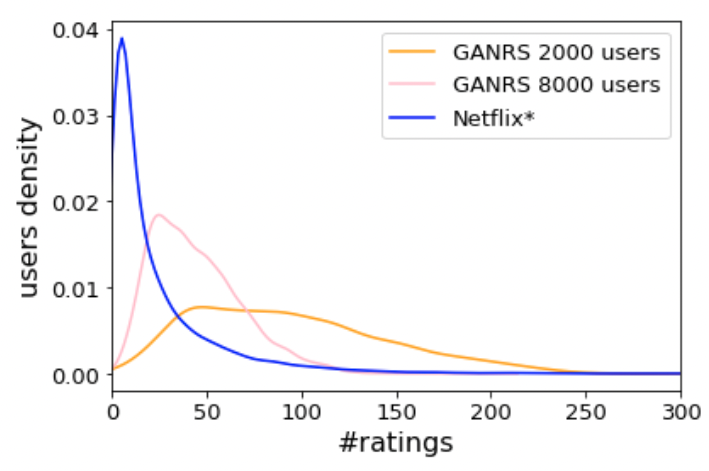}
     \caption{Distribution of users versus ratings. Number of items: 4000. Datasets 8 and 10 in Table \ref{tab:datasets}.}
     \label{fig:fig3}
\end{figure}

Figure \ref{fig:fig3} shows the density of users (y axis) that have cast different numbers of votes (x axis). (Selected datasets: 8 and 10 in Table \ref{tab:datasets}). As expected, for a fixed number of ratings in the dataset, we can observe that the higher the number of users, the lower the number of ratings. If the fixed number of samples in the dataset is distributed among a high number of users, each user centroid in the clustering stage receives a lower number of samples. Please, note that Netflix* contains 23000 users.

\subsubsection{Experiment 1b. Distribution of the user ratings}

\begin{figure}
     \centering
     \includegraphics[width=0.5\textwidth]{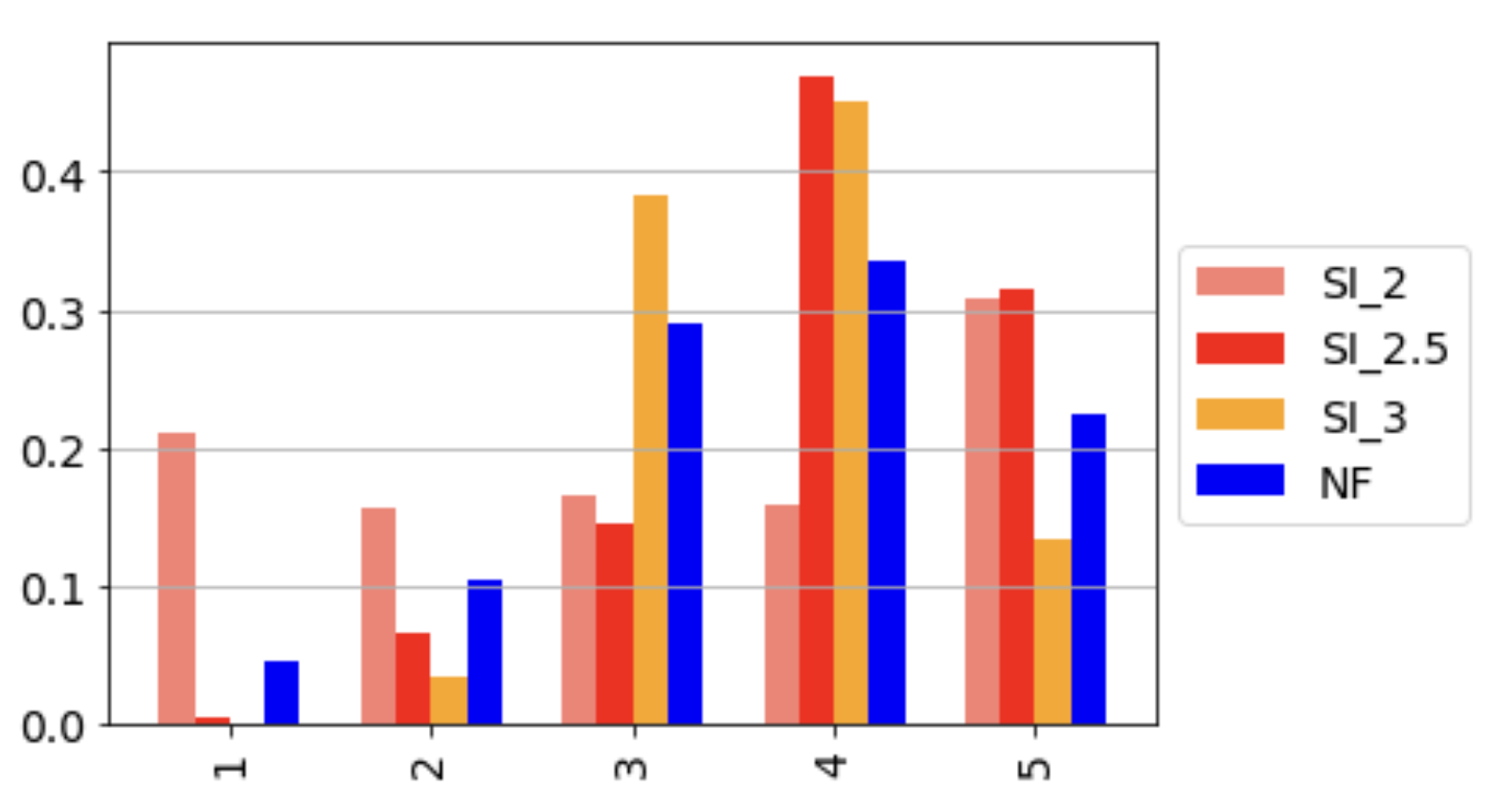}
     \caption{Distribution of user ratings. Number of users: 2000; number of items: 4000. Datasets 3, 8 and 13 in Table \ref{tab:datasets}.}
     \label{fig:fig4}
\end{figure}

Figure \ref{fig:fig4} shows the proportion of each rating {1,…,5} (x axis) when different random noise Gaussian distributions are applied. (Selected datasets: 3, 8 and 13). It can be observed that the standard deviation 2.5 generates a more similar distribution of votes, compared to the original Netflix* one, than the adjacent 2 and 2.5 standard distributions. Figure \ref{fig:fig4} also shows the impact of the Gaussian standard deviation in the layout of the individual values of the GAN generated samples.

\subsubsection{Experiment 1c. Number of repeated samples}

\begin{figure}
     \centering
     \includegraphics[width=0.5\textwidth]{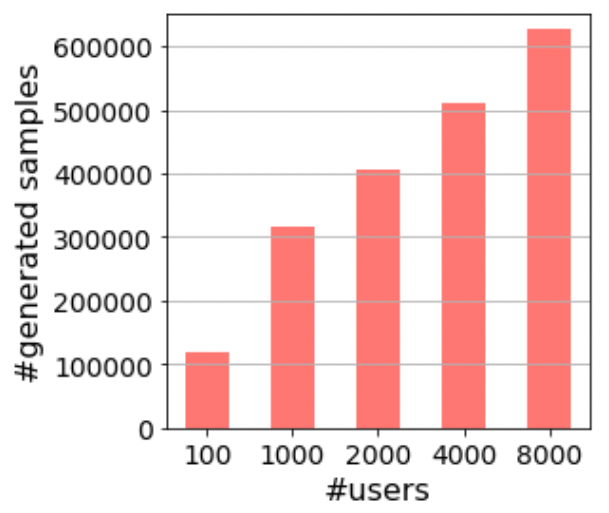}
     \caption{Number of samples remaining after removing the repeated ones. items: 4000. Datasets 6 to 10 in Table \ref{tab:datasets}.}
     \label{fig:fig5}
\end{figure}

As explained in the 'Method' section, the trained generator of the GAN predicts from random noise vectors as many dense samples as we want; all these samples are then converted from continuous dense values to discrete sparse ones. In the discretization process, repeated samples will appear that must be removed (Table \ref{tab:samplesrepresentation} contains an example). Figure \ref{fig:fig5} shows the number of samples remaining in the dataset after the removal process. The lower the number of users, the higher the number of samples assigned to each user (to its centroid in the clustering process), and then the higher the probability of repeated discrete samples. Overall, the lower the number of users, the lower the number of samples remaining. Selected datasets: 6 to 10 in Table \ref{tab:datasets}.

\subsubsection{Experiment 1d. Proportion of samples with the same user and item}
\begin{figure}
     \centering
     \includegraphics[width=0.5\textwidth]{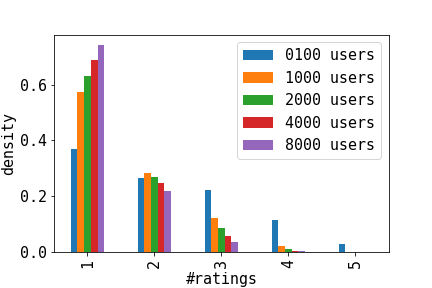}
     \caption{Proportion of samples in which the same user has cast different votes for the same item. items: 4000. Datasets 6 to 10 in Table \ref{tab:datasets}.}
     \label{fig:fig6}
\end{figure}

The GANRS generated datasets have one attribute that it does not exist in the source ones (Movielens, etc.): they contain a proportion of samples where the same user has cast different votes to the same item; e.g.: $<348, 90, 5>$, $<348, 90, 4>$, as explained in the 'Method' section. This can be seen as a mechanism to allow intermediate votes (4.5 in the example) or to allow users to change their minds. This makes sense, the number of repeated votes is two or three. The rare cases of four or five repeated votes should be better removed, as we have done in all the generated datasets. Figure \ref{fig:fig6} shows that for regular CF RS (1000 or more users), the proportion of four or five repetitions is not significant, and with the increasing of users the proportion of repetitions drops very fast.

\subsubsection{Experiment 1e.  MAE and accuracy of the dataset}

\begin{figure}
     \centering
     \includegraphics[width=\textwidth]{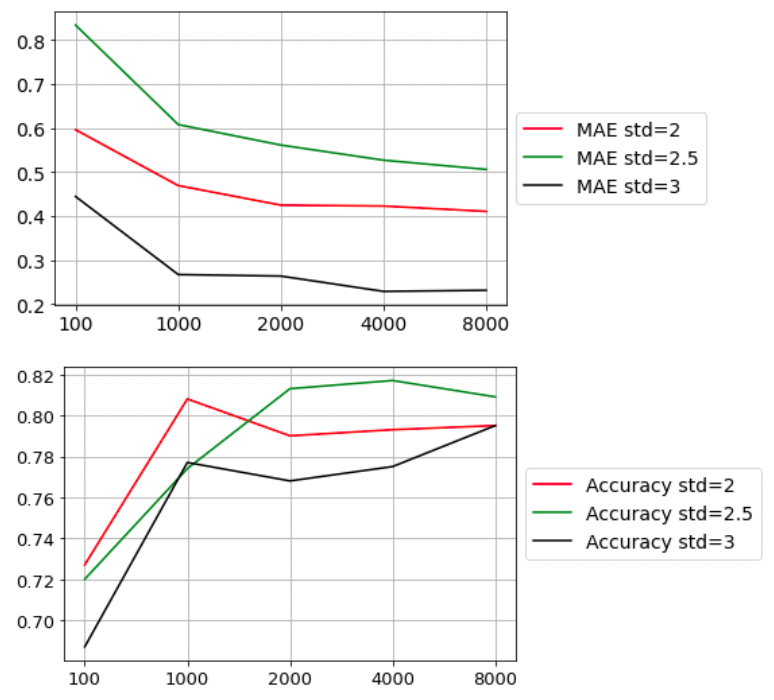}
     \caption{MAE and accuracy. Number of users: 2000; number of items: 4000; ‘std’ is the standard deviation of the Gaussian random noise distribution. Datasets 1 to 15 in Table \ref{tab:datasets}.}
     \label{fig:fig7}
\end{figure}

Whereas the previous experiments analyze the internal composition and distribution of the synthetic datasets, this experiment and the following one test the generated datasets behavior on the prediction and recommendation tasks. Figure \ref{fig:fig7} shows the prediction quality (MAE) and the accuracy of the recommendation obtained from each set of individual samples in datasets 1 to 15 of Table \ref{tab:datasets}. Please note that these measures are not obtained by analyzing and averaging the results of users. The graphs in Figure \ref{fig:fig7} show an improvement in accuracy (and its corresponding decrease in MAE error) as the number of users increases. This is the expected behavior in the CF RS, where a high number of users leads to better predictions, and it tells us that the GAN generated samples follow a CF convenient pattern. The MAE values in the top graph of Figure \ref{fig:fig7} are closely related to the distribution of ratings on each of the standard deviations 2.0, 2.5 and 3.0. MAE / accuracy results can be used to select the most appropriate standard deviation; in this case: std = 2.5.

\subsubsection{Experiment 1f. Users’ precision and recall}
\begin{figure}
     \centering
     \includegraphics[width=\textwidth]{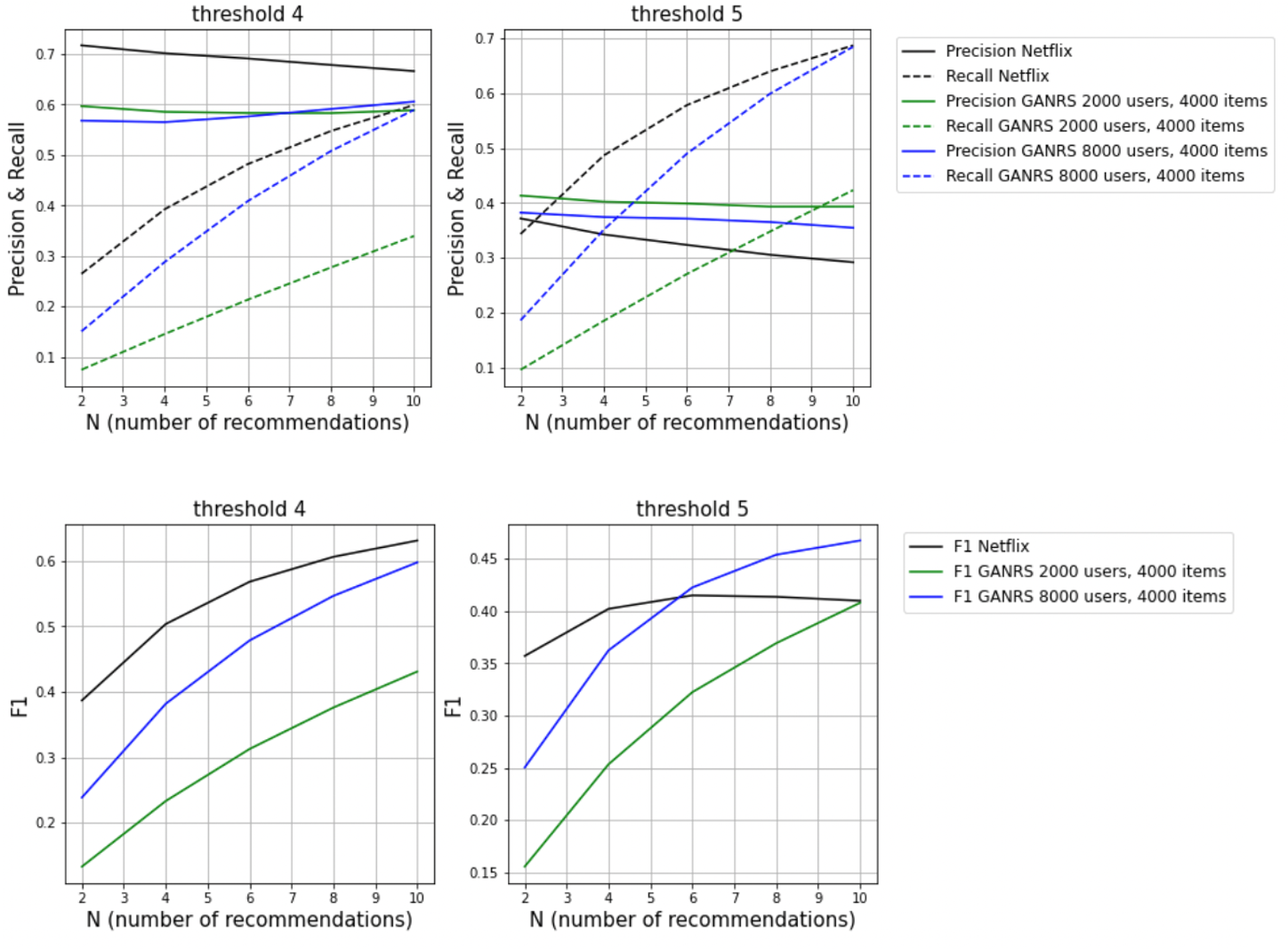}
     \caption{Precision, recall, and F1. Standard deviation of the random noise Gaussian distribution: 2.5. Number of recommendations N = [2,4,6,8,10]. Datasets 6 to 10 in Table \ref{tab:datasets}.}
     \label{fig:fig8}
\end{figure}

This experiment provides the most significant results to test out the generated datasets: we draw the values and evolutions of two representative recommendation quality measures: precision and recall. Top graphs in Figure \ref{fig:fig8} show the quality values obtained testing several numbers of recommendations N: [2,4,6,8,10] (x axis), two different relevancy thresholds $\theta$: [4,5], and two number of users: 2000 (green lines), and 8000 (blue lines). The standard deviation of the Gaussian random noise has been set to 2.5. Selected datasets: 6 to 10 in Table \ref{tab:datasets}. The values and evolutions obtained from the synthetic datasets fit with the source one: Netflix* (black lines). Additionally, as expected, the overall results of the 8000 users generated dataset improve those of the 2000 users and are more similar to the Netflix* reference (please note that Netflix* contains 23000 users). The two bottom graphs in Figure \ref{fig:fig8} represent the F1 combination of precision and recall; they clearly show the similarity in behavior of the generated datasets compared to the source one. 

\subsubsection{Experiment 2a. MAE and accuracy when the number of items varies.}
\begin{figure}
     \centering
     \includegraphics[width=0.5\textwidth]{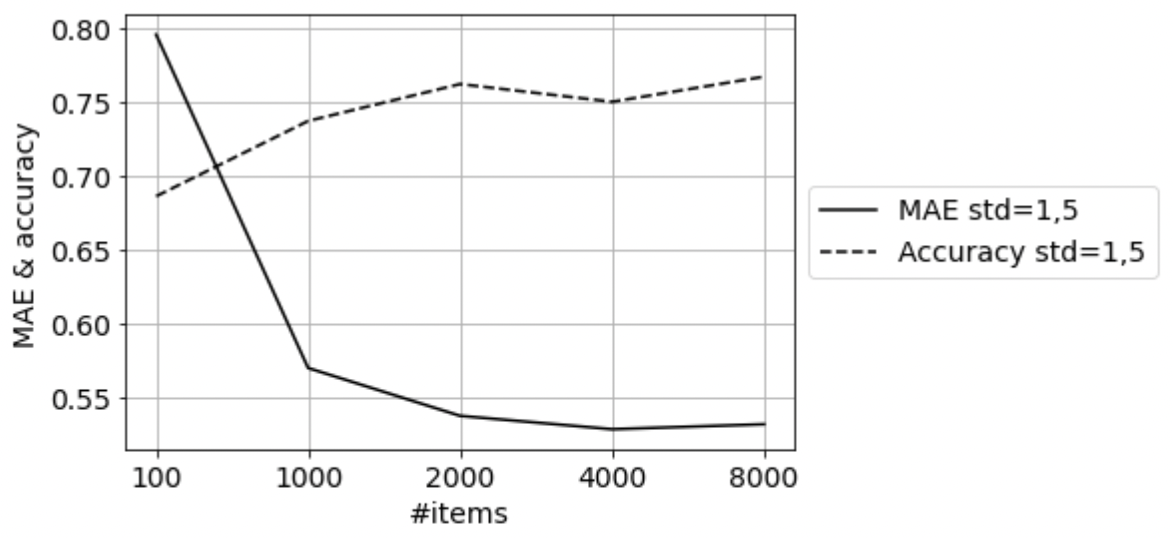}
     \caption{MAE and accuracy obtained from the dataset samples when the number of items varies. Number of users: 4000. Standard deviation of the Gaussian random noise: 1.5. Datasets 16 and 17 in Table \ref{tab:datasets}.}
     \label{fig:fig9}
\end{figure}

Experiment 1e tested MAE and accuracy quality measures on datasets with different number of users. Now we put both quality measures to the test on datasets with different numbers of items: [100, 1K, 2K, 4K, 8K]. The results in Figure \ref{fig:fig9} show adequate values for both MAE and accuracy, and consistent evolutions where accuracy increases and MAE decreases as the number of items (x axis) increases. Then, the higher the number of items, the better the accuracy: It shows that the GAN generator can enrich the data. The Netflix * source dataset contains 1,750 items and we can observe in Figure \ref{fig:fig9} how the improvement slows down around this value (x-axis). Selected datasets: 16 and 17 in Table \ref{tab:datasets}, and the 100, 1000, 4000 user versions not included in Table \ref{tab:datasets}. 

\subsubsection{Experiment 2b. Items’ precision and recall}
\begin{figure}
     \centering
     \includegraphics[width=\textwidth]{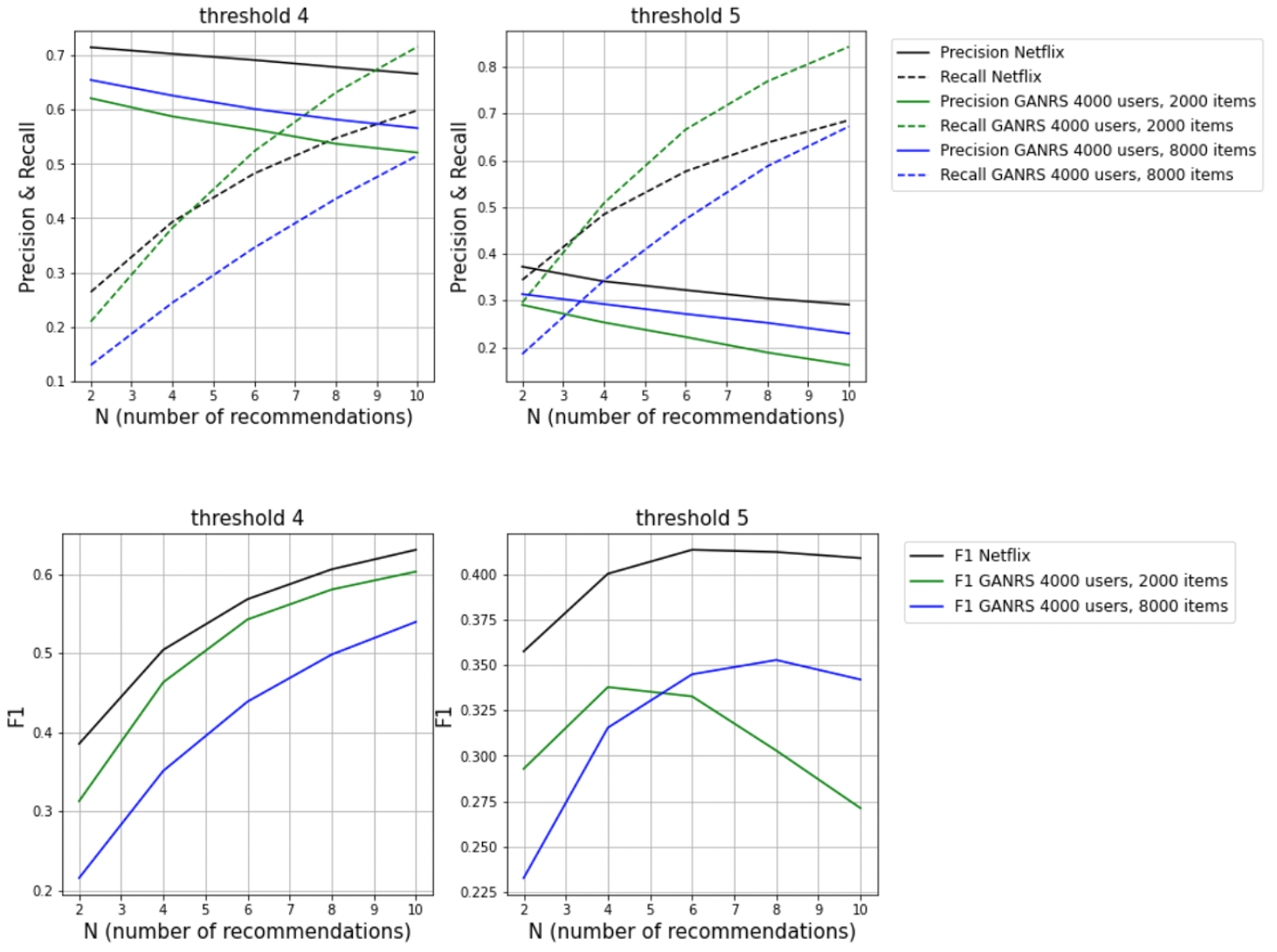}
     \caption{Precision, recall, and F1 when the number of items varies. Standard deviation of the random noise Gaussian distribution: 1.5. Number of recommendations N = [2,4,6,8,10]. Datasets 16 and 17 in Table \ref{tab:datasets}.}
     \label{fig:fig10}
\end{figure}

Experiment 2b is similar to experiment 1f; now we test the behavior of datasets that contain different number of items (instead different number of users). Figure \ref{fig:fig10} shows the performance of Netflix* (1750 items), represented using black lines, and compares it with the 2000 items dataset (green lines) and the 8000 items one (blue lines). We can observe that evolutions and values are consistent with the source datasets (black lines); furthermore, both the 2K and 4K items versions get a good performance: the first one conveniently catches the Neflix* patterns of items, since both contain a similar number of items. The second generated dataset (8K items) can enrich the data and show better accuracy than the 2K items version. Selected datasets: 16 and 17 in Table \ref{tab:datasets}, and the 100, 1000, 4000 user versions not included in Table \ref{tab:datasets}.

\subsubsection{Experiment 3a. Number of samples generated in datasets with different sizes}
\begin{figure}
     \centering
     \includegraphics[width=0.5\textwidth]{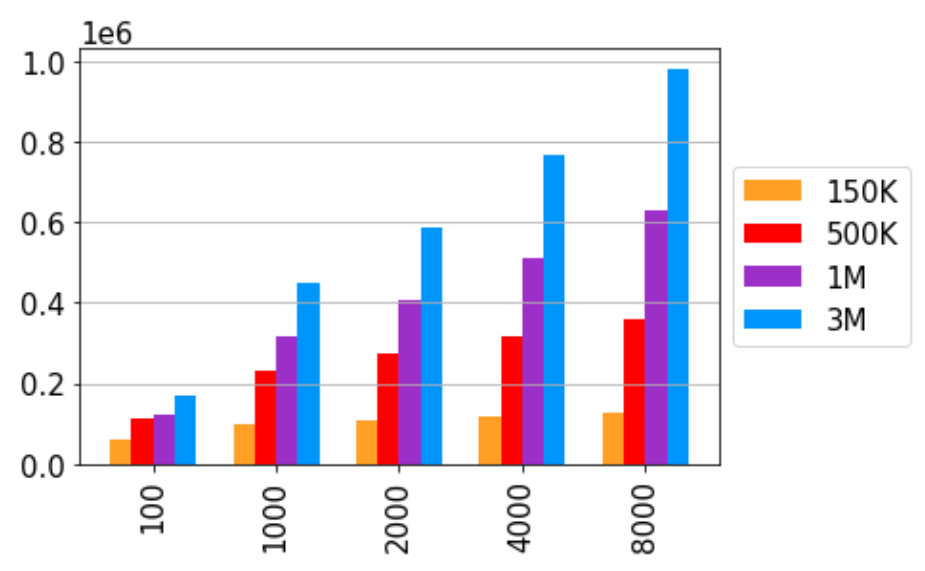}
     \caption{Number of generated samples using different number of users (x axis) and different number of GAN generated samples (legend). Standard deviation of the random noise Gaussian distribution: 1.2. Number of items: 4000. Datasets 18 to 21 in Table \ref{tab:datasets}.}
     \label{fig:fig11}
\end{figure}

Here we test the number of samples the GANRS method obtains when different numbers of generated samples and different number of users have been set. For that purpose, we define in the GAN generation process four different number of samples: 150K, 500K, 1M and 3M (datasets 18 to 21 in Table \ref{tab:datasets}, and their equivalent datasets for 100, 1000, 4000 and 8000 users). The number of items is fixed in 4K for all experiments. In figure \ref{fig:fig11} we can observe that the lower the number of users, the lower the number of generated samples; the reason is that the lower the number of users, the higher the number of samples assigned to each user (to each centroid in the clustering stage), and then the higher the probability of repeated samples that will be removed. As an example, Figure \ref{fig:fig11} shows that the 8K user dataset preserves, approximately, 1M samples from the GAN generated (version 3M), and 600K in version 1M. 

\subsubsection{Experiment 3b. Precision and recall on datasets with different sizes}

\begin{figure}
     \centering
     \includegraphics[width=0.5\textwidth]{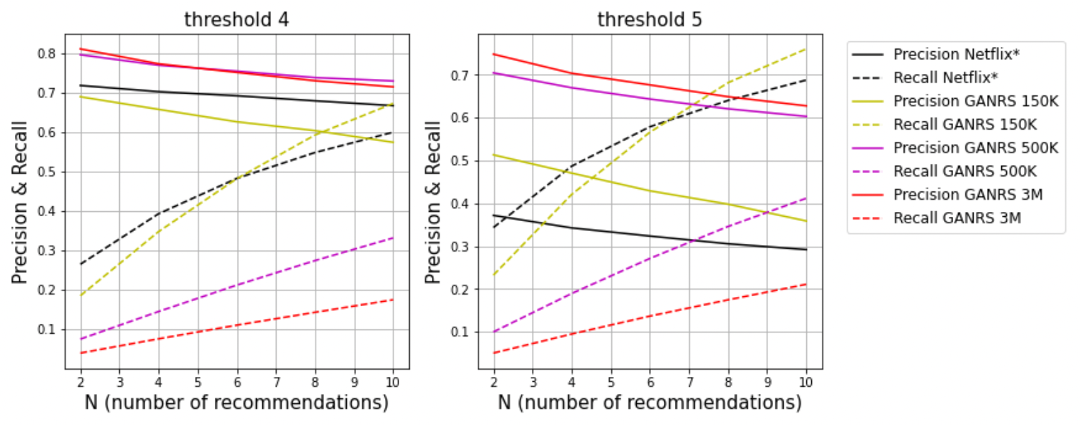}
     \caption{Precision and recall using a different number of recommendations (x axis) and a different number of GAN generated samples (legend). Standard deviation of the random noise Gaussian distribution: 1.2. Datasets 18, 19 and 21 in Table \ref{tab:datasets}.}
     \label{fig:fig12}
\end{figure}

This experiment shows the impact of increasing the number of samples on datasets with fixed parameters, in this case: 2000 users, 4000 items, and standard deviation 1.2 (Table \ref{tab:datasets}; datasets 18, 19 and 21). It is important to realize that we are using the same source dataset Netflix* to generate the three cases shown in Figure \ref{fig:fig12}: 150K samples (yellow lines), 500K samples (magenta lines), and 3M samples (red lines). Please note that 150K, 500K and 3M samples refer to the dense and continuous generated samples, before the removal stage to convert them to their sparse and discrete version. Figure \ref{fig:fig11} shows the final sizes of the datasets in the 2000 user data (x axis).

Figure \ref{fig:fig12} compares the precision and recall obtained in the Netflix* dataset (black lines) with the generated ones. Overall: 1) precision increases and recall decreases; 2) the bigger the generated dataset, the better its precision; 3) the higher the dataset, the lower its recall. Using large datasets, precision results improve since there are more relevant samples to choose, and then it is easier to succeed in the fixed number N of recommended predictions. On the other hand, recall gets worse using large datasets because they contain more variability in the samples, particularly when large standard deviations have been chosen for the random noise Gaussian distribution. Unlike precision, whose denominator is the constant N (number of recommendations), the recall quality measure depends on the variable: ‘number of relevant votes’ in the set of test items for each tested user. As the number of samples increases, the number of votes for user also increases (and from them, the number of his relevant votes); this is the reason why recall is lower in the 3M synthetic dataset of Figure \ref{fig:fig12}, and higher in the 150K version.

\begin{figure}
     \centering
     \includegraphics[width=0.7\textwidth]{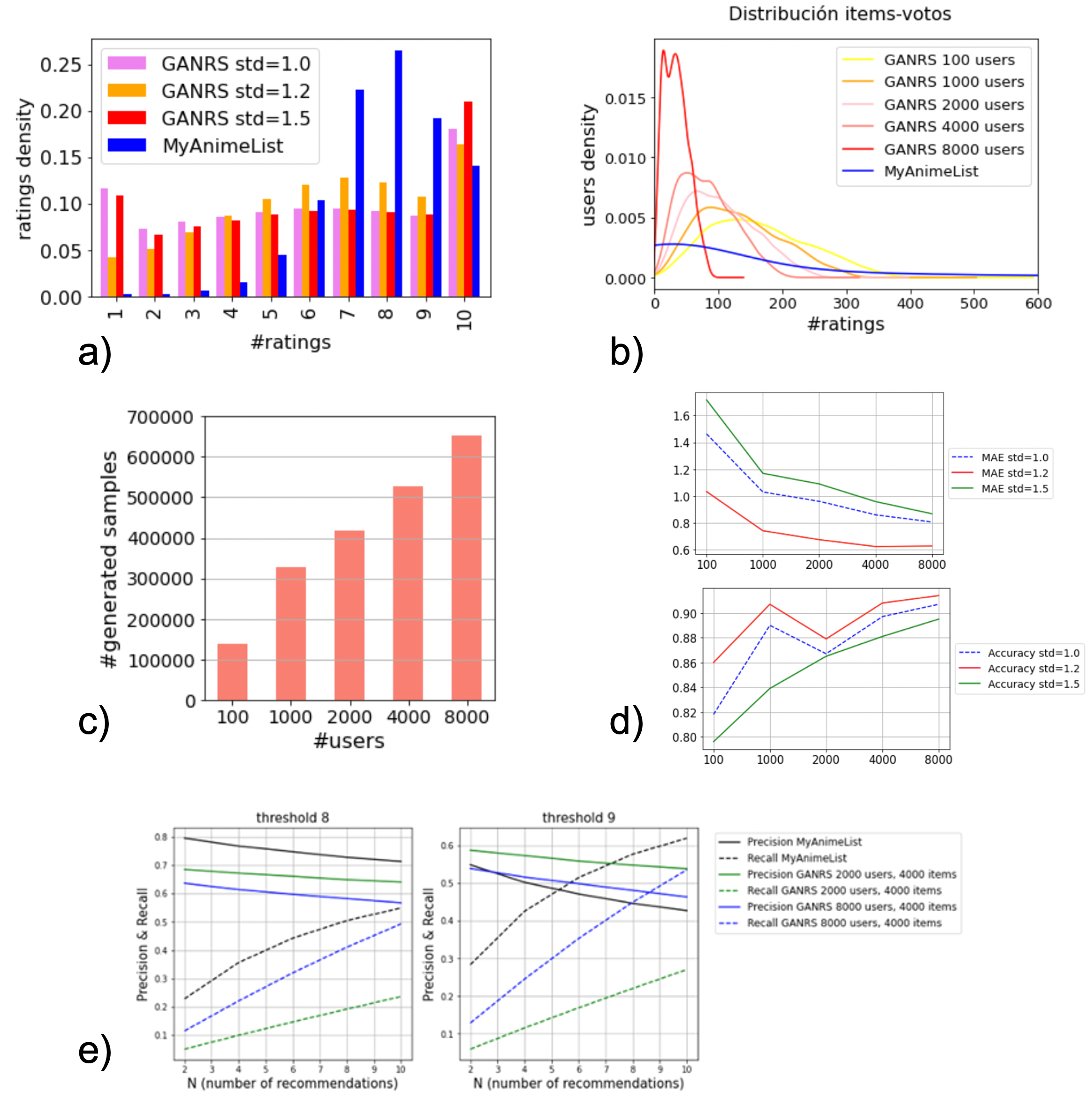}
     \caption{MyAnimeList. 1.5 million generated samples, a) distribution of the MyAnimeList ratings 1 to 10, b) distribution of users according to their number of casted ratings, c) number of samples after the removing process of the repeated ones, d) error and accuracy by processing the samples of the dataset, e) CF precision and recall (by testing the dataset users). The GANRS std=1.2 value has been set to test experiments b) to e).}
     \label{fig:fig13}
\end{figure}

\begin{figure}
     \centering
     \includegraphics[width=0.7\textwidth]{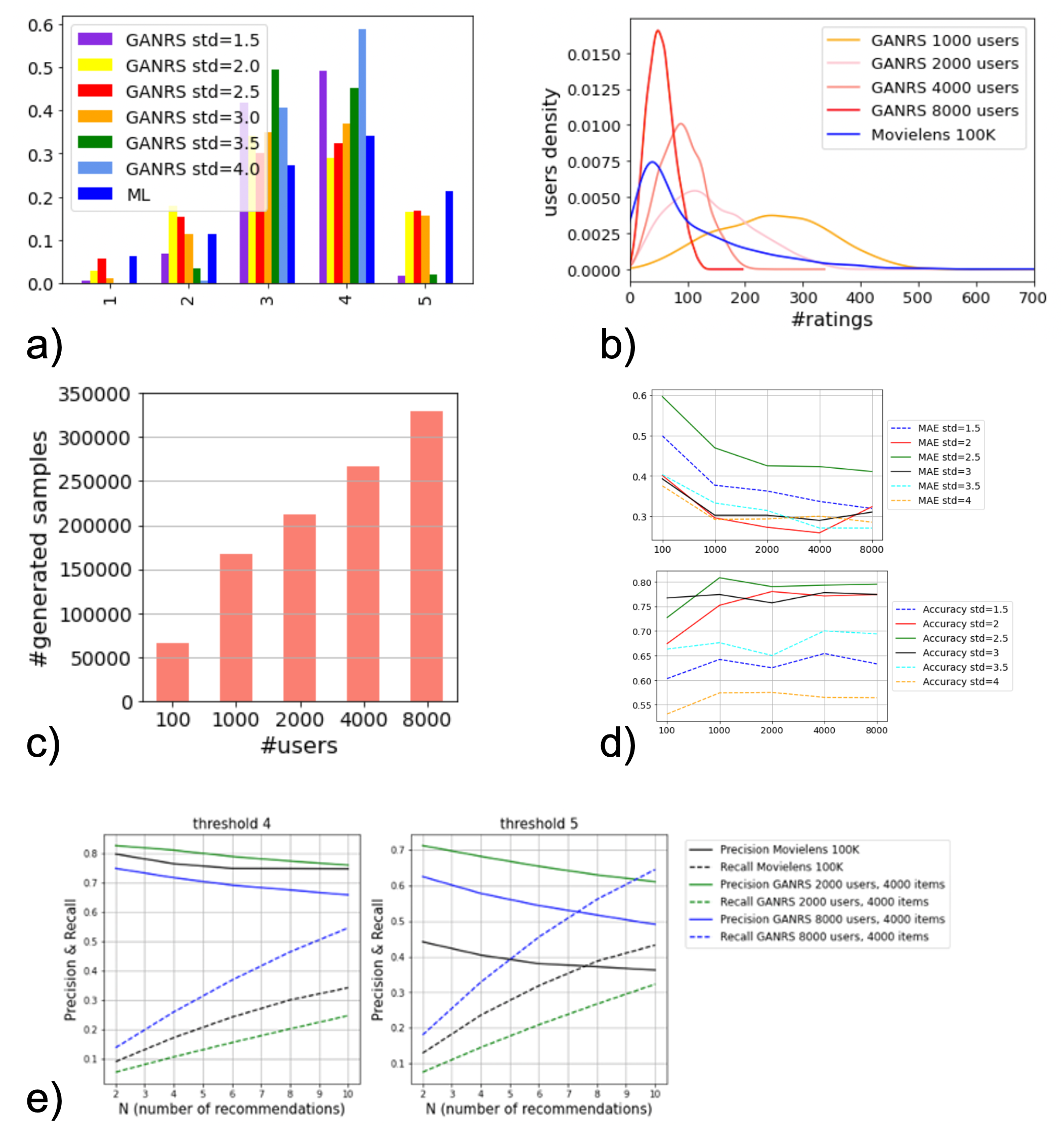}
     \caption{Movielens 100K results. 1 million generated samples, a) Distribution of the Movielens 100K ratings 1 to 5, b) Distribution of users according to their number of casted ratings, c) Number of samples after the removal process of the repeated ones, d) error and accuracy by processing the samples of the dataset, e) CF precision and recall (by testing the dataset users). The GANRS std=2.5 value has been set to test experiments b) to e).}
     \label{fig:fig14}
\end{figure}

Figures \ref{fig:fig13} and \ref{fig:fig14} show, respectively, the results obtained from the MyAnimeList and Movielens 100K test datasets. Graphs 'a)' compare the rating distribution of each source dataset (blue color) with the generated rating distributions obtained by setting different values of the Gaussian random noise standard deviation. We have chosen the standard deviation value 1.2 for MyAnimeList, and the standard deviation value 2.5 for Movielens 100K, since they obtained distributions of ratings are the closest ones to their respective baselines. Results 'b)', 'c)' and 'e)' are obtained using the selected standard deviation values. Graphs 'b)' show the distribution of users according to their number of casted ratings (x-axis). As expected, they follow the same pattern as the one in Netflix*. To compare results, please note that MyAnimelist dataset contains 19179 users, and Movielens 100K contains 943 users. Graphs 'c)' show the number of samples left after removing repeated instances. The higher the number of users, the lower the probability to generate samples containing the same user ID, item ID, and rating. In the MyAnimeList case, we started from 1.5 million generated samples, whereas for Movielens we selected 1 million generated samples. Graphs 'd)' refer to MAE error and accuracy values obtained by processing individual samples contained in each dataset. As usual in the CF context, the higher the number of users, the lower the error, and the higher the accuracy. Finally, 'e)' graphs 'e' test the recommendations obtained by processing the users in each dataset. Just as in the Netflix* case, compared to baselines, precision improves and recall gets worse.

\subsection{Discussion}

A large number of synthetic datasets have been generated to test the performance of the proposed GANRS method. These datasets have been created setting different values for the main parameters of the method: number of users, number of items, Gaussian random noise variation, and number of generated samples. To generalize the conclusions of this paper, three open and representative CF datasets have been used as sources for the generative process. Finally, a variety of quality measures have been tested on the generated datasets; from them, precision and recall are the most relevant. A key question is that we are not able to visually test the quality of the generated samples, as it can be done, for example, with the popular fake faces; indeed, in the CF context, we only can adequately test the generated datasets by comparing their CF quality results with those usually obtained in real CF datasets. For this reason, we have focused the designed experiments on those testing the precision and recall quality measures: using datasets containing different number of users, different number of items, and different number of samples (sizes). In all cases, comparatively, we obtain excellent precision results and moderate recall values. Overall, it can be considered as positive in the CF context, where precision errors are serious and recall errors are less important: It is worse to recommend a trip you will not like (sorry, no refunds!) than not to recommend a trip that you probably would enjoy. Please note that it is opposite to a deep learning model to detect malignant tumors: it is worse to make precision errors (no early detection of the tumor) than making recall errors (to erroneously detect a tumor). 

Additionally, experiments show the relevant impact of the standard deviation on the quality of the results. The GAN network learning has been based on a vector containing noise values that serve as a seed to generate the different samples on the synthetic dataset. Each ‘fake’ sample is generated from the list of random values in the ‘noise’ vector. As usual in the GAN context, random values have been created from a Gaussian distribution whose mean is 0 and its standard deviation is 1. Each generated sample contains a dense item representation, a dense user representation, and an individual value that codes the rating casted for the user to the item. Once the GAN has learned, its generative model can be used, in a feedforward process, to generate as many samples as we want, starting from a different random noise vector for each generated sample. Experimental results show that using a Gaussian distribution with standard deviation 1 leads to many ratings in the middle of the votes range: rating 3 and nearest ones in the range of votes 1 to 5 (Movielens and Netflix), and rating 5 and nearest ones in the range of votes 1 to 10 (MyAnimeList). Several experiments in this section show that we can modulate the standard deviation of the random noise Gaussian distribution to feedforward generate a wider range of ratings. As expected, when the standard deviation increases, the range of ratings also increases proportionally.

Finally, the exposed experiments include the existing relation between the number of fake samples generated for the GAN and the number of samples that will eventually contain the dataset. As explained in the ‘Method’ section, the conversion from dense and continuous values to sparse and discrete leads to a probability of sample repetitions. Results show that, as expected, the larger the number of users and items in the synthetic dataset, the lower the number of repeated samples. It has also been shown that for a usual number of users, let us say 4000 or more, the probability of more than two different ratings from a user to the same item can be considered as negligible.  

\section{Conclusions}

This paper provides an innovative method to generate synthetic parameterized collaborative filtering datasets from real ones. Synthetic datasets can be generated by selecting different numbers of users, items, samples, and distribution variability. This means that comparative experiments can be designed on the basis of a whole ‘family’ of generated datasets, for example, to test the accuracy of a new matrix factorization model when the number of users increases. A generative adversarial network is used to obtain 'fake' samples from real ones, benefitting from the inherent capacity of GAN networks to catch complex patterns in the source datasets. The GAN learns from dense and continuous embedding representations of items and users, rather than the sparse and discrete representations of the collaborative filtering datasets. The effect is a fast and accurate learning process.

The proposed GANRS method contains a clustering stage to convert from the dense generated ‘fake’ samples to the sparse and discrete values necessary to fill the generated dataset. This clustering stage implements a k-means algorithm to group items and another k-means to group users. In a natural way, both parameters 'k' set the chosen number of users and items in the dataset. A drawback of the discretization process is the generation of identical samples that our method just removes. A complete set of experiments have been made using three representative source datasets. We have tested the distribution values and evolutions of the results, as well as prediction and recommendation qualities. Although precision tends to improve, whereas recall tends to get worse, overall accuracy can be considered correct since precision is more relevant than recall in the RS context.  The results show that the generated datasets conveniently mimic the behavior of the source ones. Movielens, MyAnimeList, etc.

The source code for the proposed GANRS method is available to ensure the reproducibility of the experiments. Similarly, a complete set of generated datasets has been made available for research. This paper and its related documentation open the door to tackle some future work, such as designing alternative options to the clustering stage, to implement the PacGAN concept in the GAN discriminator, to test generated datasets using a complete range of machine learning and deep learning collaborative filtering models, to replace the GAN model for a CGAN one, generating demographically balanced datasets, and to undertake an in-depth study of the impact of the random noise vector variations in the generated set of samples. 

\textbf{\textit{Acknowledgements:}}
This work was partially supported by Ministerio de Ciencia e Innovación of Spain under the project PID2019-106493RB-I00 (DL-CEMG); the Comunidad de Madrid under Convenio Plurianual with the Universidad Politécnica de Madrid in the actuation line of Programa de Excelencia para el Profesorado Universitario; and the Plan Andaluz de Investigaci\'on, Desarrollo e Innovaci\'on (PAIDI 2020) under the project PROYEXCEL\_00257.

\bibliographystyle{unsrt}  
\bibliography{rs-bib-deep}  


\appendix Appendix A

\begin{table}[H]
\begin{footnotesize}
\begin{center}
\begin{tabular}{cc}
\hline
        \textbf{DeepMF}&\textbf{values}\\
\hline
Embedding size (both for users an items)	&5\\
Optimizer	&Adam\\
Loss function	&Mean squared error\\
Epochs &	20\\
\hline
\textbf{GAN generator} 	&\\
\hline
Input shape, noise vector size	&100\\
Block 1 dense layer \#neurons	&10\\
Block 1 activation function	& LeakyRelu, alpha 0.2\\
Block 1 normalization	&BatchNormalization, momentum 0.8\\
Block 2 dense layer \#neurons &	20\\
Block 2 activation function&	LeakyRelu, alpha 0.2\\
Block 2 regularization	&Dropout 0.2\\
Block 3 dense layer \#neurons	&$2*~embedding~size~+~
1$\\
Block 3 activation function	&linear\\
\hline
\textbf{GAN discriminator}	&\\
\hline
Input: shape 	&$2*~embedding~size~+~1$\\
Block 1 dense layer \#neurons 	&6\\
Block 1 activation function	&LeakyRelu, alpha 0.2\\
Block 2 dense layer 	&1\\
Block 2 activation function	&Sigmoid\\
\hline
\textbf{GAN train}	&\\
\hline
Epochs	&20\\
Batch size 	&64\\
Stochastic noise	&Gaussian (0,1)\\
Loss function 	&$(real~samples~loss~+~fake~samples~loss) /2$\\

\hline
\end{tabular}
\caption{Main parameter and hyperparameter values set for the neural models involved in the RSGAN method.}
\label{tab:appa}
\end{center}
\end{footnotesize}
\end{table}

\appendix Appendix B
\begin{figure}
     \centering
     \includegraphics[width=0.6\textwidth]{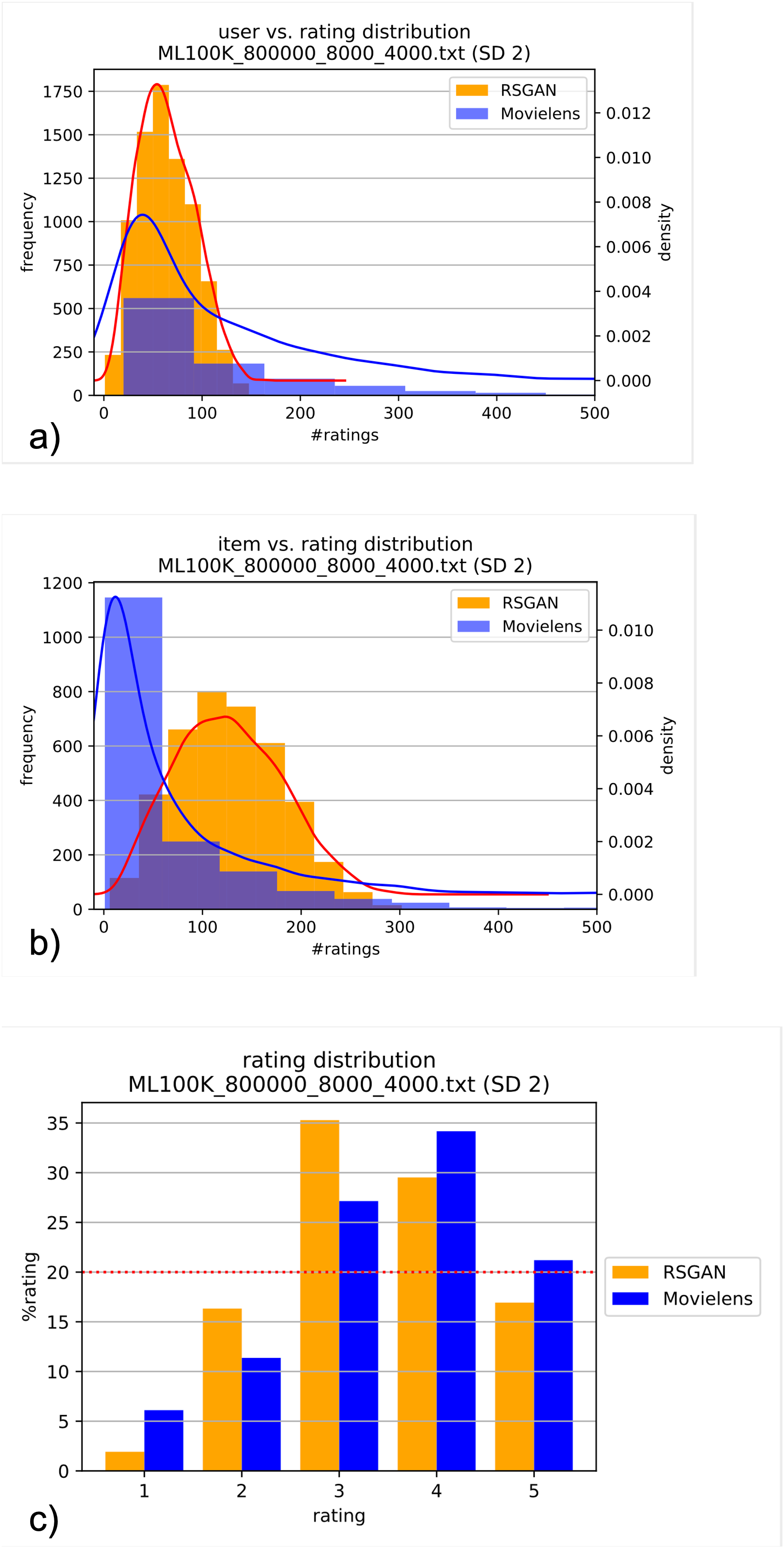}
     \caption{Main distributions of the data in the synthetic dataset generated from Movielens 100K compared to the distributions of the data in the source dataset. Number of users: 8000, number of items: 4000, initial number of samples: 800000, standard deviation of the gaussian noise: 2.5. Graph a) shows the distribution of the fake users (y axis) versus the number of ratings belonging to each of the users (x axis). Graph b) shows the distribution of the fake items (y axis) versus the number of ratings belonging to each of the items (x axis). Graph c) shows the percentage of ratings (y axis) for each of the available vote values {1, 2, 3, 4, 5} (x axis) in the dataset.}
     \label{fig:appb}
\end{figure}

\end{document}